\documentclass[times]{oupau}

\usepackage{bookmark}
\usepackage{orcidlink}
\usepackage{graphicx}
\usepackage{enumitem}
\usepackage{tikz}
\usepackage{subcaption}
\usepackage{multirow}
\usepackage{comment} 
\usepackage{relsize}
\usepackage{amsmath}
\usepackage{booktabs}

\DeclareMathOperator*{\argmin}{arg\,min}
\definecolor{lightgray}{rgb}{0.8, 0.8, 0.8}
\newcommand{\rulesep}{\unskip\ \vrule\ }

\begin{document}
\title{Enhancing web traffic attacks identification through ensemble methods and feature selection}

\author{Daniel Urda\affil{a}\orcidlink{0000-0003-2662-798X}, Branly Martínez\affil{a}, Nu{\~n}o Basurto\affil{a}\orcidlink{0000-0001-7289-4689}, Meelis Kull\affil{b}\orcidlink{0000-0001-9257-595X}, {\'A}ngel Arroyo\affil{a}\orcidlink{0000-0002-1614-9075}, {\'A}lvaro Herrero\affil{a}\orcidlink{0000-0002-2444-5384}}

\address{
\affilnum{a}Grupo de Inteligencia Computacional Aplicada (GICAP)\\  Departamento de Digitalizaci{\'o}n,  Escuela Polit{\'e}cnica Superior,  Universidad de Burgos\\
Av. Cantabria s/n, 09006,  Burgos, Spain.\\
\affilnum{b}Institute of Computer Science, University of Tartu, Estonia\\
}

\begin{abstract}
Websites, as essential digital assets, are highly vulnerable to cyberattacks because of their high traffic volume and the significant impact of breaches. This study aims to enhance the identification of web traffic attacks by leveraging machine learning techniques. A methodology was proposed to extract relevant features from HTTP traces using the CSIC2010 v2 dataset, which simulates e-commerce web traffic. Ensemble methods, such as Random Forest and Extreme Gradient Boosting, were employed and compared against baseline classifiers, including k-nearest Neighbor, LASSO, and Support Vector Machines. The results demonstrate that the ensemble methods outperform baseline classifiers by approximately 20\% in predictive accuracy, achieving an Area Under the ROC Curve (AUC) of 0.989. Feature selection methods such as Information Gain, LASSO, and Random Forest further enhance the robustness of these models. This study highlights the efficacy of ensemble models in improving attack detection while minimizing performance variability, offering a practical framework for securing web traffic in diverse application contexts.


\end{abstract}

\received{\today}
\maketitle

\section{Introduction}
\label{chap:Introduction}

During the last 25 years, digital technologies have emerged as key to shaping and amplifying disruptions across society and industries~\cite{aastrom:2022}. In today's world, there is a huge variety of digital assets associated with servers, applications, services, sensors, Internet of Things (IoT) devices, and wearables, among others, which are permanently accessible through the Internet. Although this gives people the opportunity to access a vast amount of information and data from around the world, its counterpart is that having private and exclusive networks for trusted users and collaborators is no longer a practical reality. In addition to gaining anonymous access to any resource, this widespread access exposes digital assets to various threats and vulnerabilities, such as SQL injection~\cite{SQLINJECTION-MAS} or DDoS attacks~\cite{Mirkovic:2004}, which can be exploited by malicious actors seeking to cause damage or gain unauthorized access to digital systems~\cite{enisa:2020}. These risks are particularly acute in areas such as healthcare~\cite{Esteva:2019}, bioinformatics~\cite{Urda:2018b}, industry~\cite{Thorsten:2016}, banking~\cite{alzoubi:2022}, and public sector~\cite{gao:2020}, where the consequences of a security breach can be catastrophic.

Among all digital products, websites are often one of the most targeted systems by hackers because of either the economic damage caused to companies or the personal benefits obtained from these accessed resources ~\cite{Algarni:2021}. This is even more accentuated depending on the type of data that has been compromised in the attack (e.g., personal data of customers or intellectual property). Additionally, it is necessary to consider the main targets of the attackers in order to anticipate possible security risks~\cite{Sandeep:2017}, with stealing of sensitive information or defacement being the most frequent ($64\%$ of all the attacks between both). In this sense, it is necessary to delve deeper into the vulnerabilities contained in a website, such as cross-site scripting, injection flaws, and unvalidated redirects and forwards, among others. 

Although security issues can typically be solved, or at least minimized, by system updates, this simple step is often not easy to perform in the business sector. In the case of a company website (or any digital system within the company), the company can find obstacles to maintaining its website because of domain-imposed restrictions, resistance to change, customer desires, or even developers' skills and confidence~\cite{Leppanen:2015}. In addition, a more decisive factor that may be added on top of these ones in small- or medium-sized companies is purely economic, since these companies would barely have enough economic capacity to address major changes in their digital systems without having any negative impact on their normal business operations~\cite{Radicic:2023,Gonzalez-Tamayo:2023}.

Artificial Intelligence (AI) and, in particular, Machine Learning (ML) have emerged as powerful tools for securing digital systems. By analyzing the traffic and access data generated by digital communications over time~\cite{MaganCarrion:2020,Magan:2022}, ML algorithms can identify potential intrusions and security breaches, enabling more effective intrusion detection and prevention~\cite{Gwang-Myong:2022}. Specifically, intelligent methods have been developed to monitor website vulnerability. For instance, a vulnerability scanning process of the website was developed in~\cite{Xu:2022} using log records and HTML output, helping users discover the website's vulnerability information quickly and perform actions to patch the website in time. More generally, an efficient algorithmic study and tool to detect a variety of web security vulnerabilities is capable of detecting several less common vulnerabilities, such as shell injection or file inclusion~\cite{Long:2020}. Identifying anomalies and vulnerabilities in website traffic can significantly impact network performance, quality of service provided, and overall user experience with the corresponding website application~\cite{Fernandes:2019}. Several well-known datasets, such as CSIC2010 v2~\cite{CSIC2010} and CSE-CIC-IDS2018~\cite{sharafaldin:2018} have been used to develop AI/ML-based solutions that can detect specific outcomes of interest. The CSIC-2010 v2 dataset comprises thousands of web requests automatically generated for the application of an e-commerce website by the ``Information Security Institute'' of the Spanish Research National Council.

This study aims to investigate the development of AI/ML-based solutions that can accurately identify attacks in website traffic. For this purpose, the CSIC2010 v2 dataset~\cite{CSIC2010} was employed to develop several classifiers and test their predictive performance, because this dataset is a well-known web traffic dataset for e-commerce applications. Furthermore, this work describes a feature extraction procedure that can be applied to this and other similar web application contexts to extract rich information encoded in HTTP traces, thus preprocessing this dataset and making it suitable for training ML models. Additionally, this study analyzes the benefit of using ensemble classifiers in terms of predictive performance, such as Random Forest (RF) and Extreme Gradient Boosting (XGBoost), rather than traditional classifiers, such as k-nearest neighbor (kNN), LASSO and Support Vector Machines (SVM). In addition, the impact of using feature selection methods such as Information Gain (IG), LASSO, and RF in combination with classifiers is also analyzed because less complex classifiers are commonly preferred when the collection of variables required by the models comes with some difficulties and possibly additional costs.

The remainder of this paper is organized as follows. Section~\ref{sec:RelatedWork} includes a literature review of the most relevant and recent studies in this field. Sections~\ref{sec:Dataset} and~\ref{sec:Methods} describe the dataset and the variety of methods used in this study to develop predictive models. The experimental design used in the analysis is described in Section~\ref{sec:ExperimentalDesign} and the results obtained are presented in Section~\ref{sec:Results}. Finally, Section~\ref{sec:ConcAndFW} provides the conclusions derived from this study and some guidelines for future work.
\section{Related work}
\label{sec:RelatedWork}

Unsupervised learning techniques have previously been employed to detect anomalous behavior or attacks in the website context. In~\cite{Stevanovic:2013}, Self-Organizing Maps (SOMs) and Modified Adaptive Resonance Theory 2 (Modified ART2) were used to gain more knowledge about the distribution and visitor profiles of websites, discovering a remarkable difference between malicious web crawlers and other visitor groups. In addition, the authors highlighted that more than half of the malicious web crawlers browsed the Internet in a very similar way to humans. More recently, a deep learning approach has achieved $98\%$ accuracy when classifying attacks based on web requests~\cite{Pillai:2023}. The deep learning approach presented by the authors in this work consisted of two stages: a first stage based on denoising and stacked autoencoders to distinguish between normal and anomalous web requests, and the second stage, which feeds the anomalous web requests identified to supervise learning models that classify the anomaly. Focusing on the CSIC2010 v2 dataset, the authors of ~\cite{Atienza:2015} employed projection models, such as Principal Component Analysis (PCA) or Cooperative Maximum Likelihood Hebbian Learning (CMLHL), together with SOMs to provide new insights into anomalous situations in HTTP traffic. They concluded that there were significant challenges to overcome in the application of models for the ability to discriminate between different existing categories of instances.

Several relevant studies have applied supervised learning techniques to classify network-related patterns. The authors in~\cite{Gniewkowski:2021} trained several ML models, such as RF, employing features extracted from an unsupervised language representation model to embed HTTP requests, and obtained state-of-the-art accuracy in three different datasets. In~\cite{Rong:2022}, the authors used neural network learning to predict the number of visits to web pages, which allows the network manager to adjust the scheduling strategy to guarantee user experience. Other researchers~\cite{Lian:2020} have relied on Convolutional Neural Networks (CNN) to extract semantic features automatically to further use them with a well-known SVM classifier to achieve state-of-the-art performance metrics compared to other deep learning-based approaches. More recently, the CSIC2010 v2 dataset was used in~\cite{Urda:2023} to generate different instances differing in the features extracted within the web request, performing an in-depth analysis where well-known classifiers such as LASSO, \emph{k}NN, or SVM were used to predict normal and anomalous traffic. The kernel-based classifier (SVM) outperformed the other two algorithms, achieving an AUC of $0.83$.

Ensemble methods have been shown to provide high performance rates in several domains in contrast to traditional ML methods. In the ensemble paradigm, several base learners are trained to solve a given problem, whereas the output of the ensemble combines the output of every base learner in a certain manner. In the context of attack identification, the authors in~\cite{Kamarudin:2017} proposed a system that combined a filtering method to remove irrelevant features with LogitBoost using an RF as a weak classifier. They tested this approach on a well-known netflow dataset, UNSW-NB15, which achieved very good values of low false rejections. More recently, an ensemble system called M-AdaBoost-A was published in~\cite{Zhou:2020} incorporated the AUC metric within the boosting system in order to allow dealing with imbalance problems as commonly occurs within intrusion detection. Furthermore, the PSO-XGBoost algorithm employs swarm particles with other ensemble methods, such as AdaBoost or RF, to develop an accurate classifier that was validated in a well-known netflow dataset (NLS-KDD). Additionally, the authors in~\cite{Zivkovic:2022} presented an XGBoost model tuned with a novel hybrid firefly algorithm that improves the classification accuracy and average precision of network intrusion detection systems. Their proposed method was first validated in 28 well-known CEC2013 benchmark instances and was later tested on the NSL-KDD and USNW-NB15 datasets.

Additionally, ensembles have been employed for web attack detection in the Internet of Things (IoT), a context that considers the existing diversity in these devices and results in a more complex interaction environment. The authors in~\cite{Luo:2021} used three different deep learning models to detect attacks separately (Modified Random Network, Convolutional Neural Network and Long-Short Term Memory network) and finally employed a Multi-Layer Perceptron to combine the intermediate vectors as the final ensemble classifier. Using this approach, they outperformed individual deep-learning models on a real-world dataset collected by a security company. Furthermore, the advanced deep learning-based approach described in~\cite{Saharkhizan:2020} was implemented to detect attacks against IoT systems by integrating a set of Long Short-Term Memory modules into an ensemble classifier consisting of a decision tree that provides an aggregated output. In this study, the authors were able to achieve accuracy rates of over $99\%$ using a real-world dataset of Modbus network traffic.


Although deep learning methods such as Convolutional Neural Networks and Recurrent Neural Networks have demonstrated exceptional performance in various cybersecurity applications, their implementation often entails substantial computational costs and extensive hyperparameter tuning. In addition, a wide variety of methods in the above-described works have already been employed over netflow datasets or IoT environments, although the authors of this study identified that ensemble methods have not been fully explored to detect attacks on web requests. Although existing studies have demonstrated the advantages of employing AI/ML-based methodologies for the accurate prediction of normal and anomalous traffic in website contexts, to the best of our knowledge, the development of ensemble methods within the CSIC2010 v2 dataset has not been studied previously. While deep learning methods are pivotal in many domains, particularly in areas such as image recognition and natural language processing, this study focuses on exploring the potential of ensemble methods in this dataset, which belongs to a very different context that significantly differs from those where deep learning has traditionally excelled. Moreover, few studies employ feature selection (FS) methods to remove irrelevant features that may introduce noise to the classifiers. This work also addresses the need to analyze the impact of FS methods on classifiers trained to identify attacks in web requests.

In addition, the authors of this work assess that many of the features of this dataset are underutilized, such as URL requests, where some information has not been appropriately exploited in the past. At the same time, existing studies lack a comprehensive elucidation of the treatment of these features to feed and train ML-based classifiers. Therefore, the authors aimed to describe an approach to preprocess the features of the original CSIC2010 v2 dataset and extract new features that may enhance the predictive performance of ensemble-based classifiers. Furthermore, the proposed preprocessing procedure should be applicable to websites with similar HTTP request compositions.

Overall, this study attempts to fill the gaps previously mentioned, and aims to generate models with greater adaptability to the intrusion detection paradigm, focusing on the enhancement of attack identification in web traffic  using ensemble methods and feature selection based on the CSIC2010 v2 dataset.
\section{Dataset}
\label{sec:Dataset}

In 2010, the Spanish National Research Council (CSIC) developed the dataset used in this study, CSIC2010 v2, to develop predictive models that can accurately predict web traffic attacks. In particular, this dataset incorporates a collection of different attacks produced by HTTP queries. Furthermore, attack simulations are performed within a web application where different purchases are carried out, which could be considered a common e-commerce scenario in which users accumulate products in the shopping cart and provide certain personal data to perform the purchase. In the first row of Table~\ref{tab:dataset}, summary statistics of this ``Original'' dataset can be observed, where there are a total of $223,585$ samples, each consisting of 18 describing features, which are distributed among $104,000$ of normal traffic and $119,585$ attacks.

\begin{table*}[!t]
    \centering
    \renewcommand{\arraystretch}{1.4}
    \caption{Statistics of the original dataset and the processed one used to develop predictive models.} 
    \label{tab:dataset}
    \begin{tabular}{|c|c|c|c|c|}
    \toprule
      \textbf{CSIC-2010} & \textbf{Instances}   & \textbf{Variables} & \textbf{Normal traffic} & \textbf{Attack} \\
    \midrule
    \textbf{Original} & 223585 & 18 & 104000 & 119585 \\
    \hline
    \textbf{Processed} & 13569 & 78 & 4303 & 9266 \\
    \bottomrule
\end{tabular}
\end{table*}

To provide a dataset from which machine learning methods can learn hidden relationships, the original dataset was preprocessed by grouping samples belonging to the same session. This step was performed using the \emph{cookie} variable within the original dataset. A total of $13,569$ processed samples are obtained through this procedure, which additionally allows the entire payload of the URL request to be composed by joining the value of the \emph{payload} variable of samples belonging to the same session. This full payload is stored for further pre-processing, together with other variables of the dataset, to extract possible relevant features that may allow discriminating normal from anomalous traffic. For this purpose, a detailed description of the preprocessing procedure performed on the main features of the original dataset is provided below.

\begin{itemize}

    \item Variable ``\textbf{method}''. It contains the HTTP method of the given sample allowing one of the following values:``PUT'', ``POST'', and ``GET''. To make this information suitable for machine learning models, the well-known one-hot-encoding method was applied to generate three new binary variables that encode this information.

    \item Variable ``\textbf{payload}''. This feature includes the data transmitted in an HTTP request for the given purchase being made by the user, that is, it contains bidirectional data between the client and the web server. In general, this feature can take a string value consisting of a conjunction of zero, one, or more \emph{Key=Value} pairs represented by the operator ``\&'' (similar to the logical operator \emph{and} in any programming language). By analyzing the information contained in the payload of all the samples of the dataset, a total of $19$ different key labels are identified (e.g., ``login'', ``password'',``email'', etc.). Consequently, $19$ new binary features are extracted and included in the processed dataset in such a way that the key labels appearing within the payload are coded as 1 and 0 otherwise. With respect to the value assigned to each key label within the payload, $19$ additional features are extracted and included in the processed dataset encoding the string length of the specific value or zero when its associated key label does not appear within the full payload. Therefore, $38$ new features capture the full payload information. Furthermore, two more features that summarize the information of the $38$ new variables were added to provide as much information as possible to the models: ``num.keys'', which reflects the number of key labels that appear in the given payload, and ``total.length'', which represents the sum of the length of all the values present within the full payload. An example is illustrated in Figure~\ref{fig:payload}. In this case, the full payload contains three \emph{Key=Value} pairs, where the key labels present correspond to ``provincia'', ``B2'' and ``cantidad''. This automatically sets to 1 the corresponding key labels binary features (i.e., ``key.provincia'', ``key.B2'', ``key.cantidad'') where the remaining $16$ features are left to 0 (not represented in the figure due to space limitation). Similarly, the length of string values for each of those keys is computed and stored in their respective length feature, i.e., the length of ``Zaragoza'' is equals to 8 and is stored in ``length.provincia'', and similarly with the other two values (the other $16$ variables associated to the length of the values are left to 0, not being shown in the figure due to space limitation). And finally, ``num.keys'' and ``total.length'' contain the total of key labels and the sum of the length of the values present in the full payload, respectively.

    \begin{figure}
        \centering
        \includegraphics[width=0.9\linewidth]{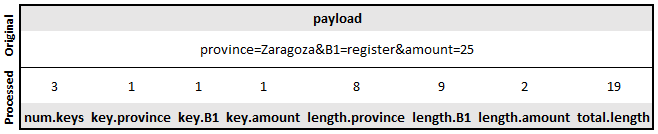}
        \caption{Example of the feature extraction procedure applied to a given payload.}
        \label{fig:payload}
    \end{figure}

    \item Variable ``\textbf{url}'': It contains the URL and resource accessed through HTTP. In order to apply a similar feature extraction procedure to the one described for the payload, it was observed that the URL always starts with the address ``http://localhost:8080/'' followed by zero, one or more directories, and finally ending with the resource file accessed. The directory structure is a common string in which the symbol ``/'' is used to separate each directory from the next one. With respect to the resource file accessed, it also includes the type of file, or file extension, at the end (e.g., ``png'', ``old'' or mostly ``jsp''). After analyzing the URL values for all samples within the original dataset, $24$ different file extensions were found. Consequently, $24$ new binary features are extracted and included in the processed dataset in such a way that the types of files accessed in the given URL are coded as 1 and 0 otherwise. Additionally, four more variables were extracted and included in the processed dataset to provide as much information as possible to the machine learning models. The first defines whether the URL value is valid (when a resource file is accessed within the local host address) or not, named ``isValidURL.” Whenever the URL value is not valid, all other features extracted from this variable will remain at 0. Next, the new variable ``numDir'' contains the number of directories included in the path. Third, the new variable ``lengthDir'' sums up the length of each individual directory in the path. Finally, a fourth new variable ``lengthFile'' includes the length of the filename accessed in the URL value. In Figure~\ref{fig:url}, an example of a valid URL with two directories in the path and a ``jsp'' file accessed is shown (the remaining 23 binary variables for the other file extensions are not shown because of space limitations, although they are set to 0). Besides, the length of the two directories within the path (``tienda1'' and ``miembros'') is set to 15 and the length of the file accessed (without the file extension) is 6 for this specific URL value.

    \begin{figure}
        \centering
        \includegraphics[width=0.6\linewidth]{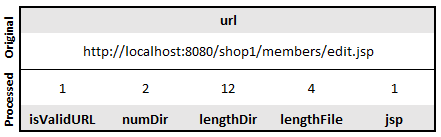}
        \caption{Example of the feature extraction procedure applied to a given URL.}
        \label{fig:url}
    \end{figure}
    
\end{itemize}

Finally, seven more variables were retained from the original dataset and added next to the $71$ extracted features through the above-mentioned procedure. Therefore, the processed dataset consists of $13,569$ samples, each of them described by $78$ input features and the binary output feature. The second row in Table~\ref{tab:dataset} shows the class distribution resulting in an imbalanced dataset with slightly more than 2x cases (attacks in this context) than controls. The increased prevalence of anomaly samples can be attributed to a higher likelihood of encountering anomalies in smaller, more isolated sessions compared with regular traffic. In this scenario, the performance metrics employed to assess the goodness of the predictive models that will be developed need to be suitable for imbalance problems, and will be described in further sections.

Although the CSIC2010 v2 dataset provides a robust benchmark for evaluating web traffic and intrusion detection systems in this study, it is worth noting its limitations. Notably, it only includes HTTP traffic and lacks HTTPS requests, which are becoming increasingly prevalent in modern web applications. Additionally, because the dataset was developed in 2010, some attack patterns and web behaviors may no longer reflect contemporary scenarios. These limitations highlight the need for future research to address encrypted traffic contexts and more up-to-date vulnerabilities, although this is beyond the scope of this study.
\section{Methods}
\label{sec:Methods}

This study considered several well-known feature selection methods and classifiers. For the former, Information Gain (IG), the LASSO, and Random Forest (RF) were used to remove irrelevant features before training any given classifier, which is motivated by the recent application of these methods in related problems~\cite{Hassani:2021,Magan:2022}. With respect to the classifiers used in this study, the authors considered three well-known classifiers as baseline methods and two more classifiers as ensemble methods to analyze the benefits of these previous methods in terms of predictive performance. The LASSO, k-Nearest Neighbor (kNN), and Support Vector Machines (SVM) were chosen as baseline classifiers because they represent different families of machine learning models (linear, distance-based, and kernel-based) that approach a given problem from different perspectives. Following a similar strategy, Random Forest and Extreme Gradient Boosting (XGBoost) were the two ensemble methods chosen, being a good representation of different families of ensemble methods such as bagging and boosting methods. Because some of the methods are used either as feature selection methods or classifiers, they will be described in more detail in the section dedicated to the classifiers.

\subsection{Feature selection methods}
As some input features may be irrelevant and introduce noise to classifiers, this study analyzed the impact on the predictive performance of using three different feature selection methods prior to training any given classifier. Next, a detailed description of the Information Gain method is provided, whereas the other two ( LASSO and Random Forest) are described in Sections~\ref{sec:baseline} and~\ref{sec:ensemble}, respectively.

\begin{itemize}
    
    \item The Information Gain (IG) algorithm~\cite{azhagusundari2013} focuses on reducing entropy as part of the dataset transformation process. In this study, IG was employed as a feature selection method to assess the gain associated with each variable, denoted as \textbf{${X_j}$} in relation to the class label, denoted as \textit{y}. This assessment is based on the entropy values, \emph{H(·)}, calculated as shown in Equation~\ref{eq:infgain}:

    \begin{equation}
    \label{eq:infgain}
        IG(y, X_j) = H(y) + H(X_j) - H(y, X_j)
    \end{equation}

\end{itemize}

\subsection{Classifiers}
\label{sec:classifiers}

This section describes the five different binary classifiers analyzed in this study to predict whether a sample corresponding to an HTTP request is normal or anomalous traffic on a website. Next, three baseline models and two ensemble models are described in detail.

\subsubsection{Baseline models}
\label{sec:baseline}

\begin{itemize}
    \item The Least Absolute Shrinkage and Selection Operator (LASSO) is a linear model in which the outcome is modeled as a linear combination of \emph{p} input features~\cite{bishop:2006}, as depicted in Equation~\ref{eq:linear_assumption}.

    \begin{equation}
    \label{eq:linear_assumption}
        \hat{y} = f(\beta_0 + \sum_{j=1}^{p}{\beta_jx_j})
    \end{equation}

    For this particular binary classification problem, $f(\cdot)$ is a logistic function that transforms any given input into the $[0,1]$ range and $\beta$ is the parameter vector that the model needs to learn by solving the minimization problem depicted in Equation~\ref{eq:lasso}.

    \begin{equation}
    \label{eq:lasso}
        \hat{\beta}_{\lambda} = \argmin_{\beta} \enspace ||y-f(\beta X^T)||_2^2 + \lambda||\beta||_1 \enspace ,
    \end{equation}

    The $L_1$-penalty that LASSO adds in the second term of Equation~\ref{eq:lasso} helps avoid overfitting by removing (i.e., setting to zero) as many $\beta$ coefficients as possible without decreasing the performance of the model. To control the strength of the regularization performed by LASSO, the $\lambda$ hyperparameter was tuned through nested cross-validation. Therefore, LASSO can also be used as a feature selection method, where only variables with their associated $\beta$ coefficients differing from zero are relevant and further used to train any machine learning model. In this study, the LASSO model was used as both a feature selection method and a binary classifier.
    
    \item The k-Nearest Neighbor (kNN) algorithm has the ability to identify intrusion attacks by measuring the distance between various instances. This is achieved by computing the local density of a test element $X_i$ through the creation of a hypersphere encompassing its \emph{k}-th nearest neighbors, where $k$ is a predefined value. Subsequently, if the computed density falls below a predefined threshold, an anomaly is flagged; conversely, if the density is high, it is classified as belonging to a target set~\cite{Sukchotrat2008}. The output of the kNN algorithm is the most frequent label of the k-nearest neighbors to the \emph{i}-th sample, measured using a distance function $d(x_i, x_j), \forall j \neq i$.
    
    \item The Support Vector Machines (SVM) stands out as one of the most extensively utilized classifiers~\cite{cortes1995support}. This machine learning model endeavors to discern a hyperplane capable of maximizing the margins of separation among distinct classes present within the training data, allowing the generalization and accurate classification of new unseen instances. Notably, this model exhibits a high sensitivity to alterations in the utilized kernel, its associated hyperparameters, and cost. The Radial Basis Function (RBF) kernel function was chosen in this study because of its effectiveness across a myriad of problem domains and its simplicity as it is governed by a sole hyper-parameter, $\gamma$. This kernel function is defined by Equation~\ref{eq:svmrbf}:

    \begin{equation}
    \label{eq:svmrbf}
    K(X,X') = exp(-\gamma||X-X'||^2)
    \end{equation}

\end{itemize}

\subsubsection{Ensemble methods}
\label{sec:ensemble}

Ensemble methods~\cite{Dietterich:2000} are statistical and computational learning techniques designed to create a collection of classifiers, often referred to as ``experts''. These methods involve the calculation of new data points, each assigned certain weights, based on the collective input or votes of various experts.

\begin{itemize}
    \item Random Forest (RF): This is a bagging algorithm based on ensembles of $N_{\text{trees}}$ decision trees~\cite{Cutler:2012}, each of them trained on a subset of samples and random variables. Given a \textit{q}-dimensional random vector represented as $X= (X_1, ...,X_q)$, which is responsible for representing the actual response value, where an unknown joint distribution is assumed. The main goal is to determine a function $f(X)$ that allows the prediction of \textit{Y}. This function is determined by a loss function $L(Y,f(X)$ and is defined such that the expected value of the loss is minimized. For a new sample, the simplest RF output averages the predictions provided by each individual decision tree. However, because each decision tree is built on a subset of samples, out-of-bag samples can be utilized to compute the performance of the given tree, $\omega_i$, and consequently, weigh its prediction according to its performance, as shown in Equation~\ref{eq:rf}.

    \begin{equation}
    \label{eq:rf}
    \hat{y} = \frac{1}{N_{\text{trees}}} \sum_{i=1}^{N_{\text{trees}}} \omega_if_i(x)
    \end{equation}

    Furthermore, the RF algorithm is typically employed beyond classification. It is also an algorithm capable of identifying relevant features~\cite{Reif:2006} by measuring the importance of each individual feature, allowing us to rank and select from categorical and continuous data, where there may be some interaction between different types of data. 
    
    \item Extreme Gradient Boosting (XGBoost): This novel boosting algorithm~\cite{Chen:2016} uses $N_{\text{learners}}$ decision trees as baseline classifiers that are optimized as more trees are added. In this sense, this algorithm has the ability to build very deep trees that allow learning of more complex relationships in the data. Likewise, new trees are added in stages in such a way that they optimize the performance of the previous ones by taking into account the residuals. This process is achieved through Gradient Boosting, which focuses on misclassified instances to improve the performance of the next tree and learns the contribution of each tree to the final model. The output of an XGBoost model is the sum of the outputs of all its baseline learners as depicted in Equation~\ref{eq:xgboost}

    \begin{equation}
    \label{eq:xgboost}
    \hat{y} = \sum_{i=1}^{N_{\text{learners}}} f_i(x)
    \end{equation}

\end{itemize}
\section{Experimental design}
\label{sec:ExperimentalDesign}

A stratified 10-fold cross-validation strategy was used to train and evaluate the performance of the different settings (i.e., feature selection method and classifier) considered in this study. This validation strategy splits the data into 10 folds of equal size such that the original distribution of the classes is approximately the same within each fold. Classifiers were trained using nine out of the ten folds (training set) and then evaluated on samples of the outer fold, commonly known as the test set. This procedure was repeated by iterating and shifting the test set, which consequently changed the training set. As a result of this validation strategy, one can calculate and estimate the average performance of a given setting in new unseen data. This robust validation strategy minimizes the potential biases introduced by the train/test split and ensures that the class distribution is preserved across the different training and test folds, providing a reliable estimation of model performance on unseen data~\cite{Xu:2018}.

To evaluate the goodness-of-fit of the trained classifiers, this study used several well-known metrics suitable for imbalanced binary classification tasks, such as \emph{Accuracy}, \emph{Precision}, \emph{Recall}, \emph{F1-score}, \emph{Gmean}, and \emph{AUC}. True Negatives (TN), True Positives (TP), False Negatives (FN), and False Positives (FP) are calculated on the test set by thresholding the output probability vector and comparing the resulting output label to the ground truth (typically, probabilities equal to or greater than 0.5 are mapped to the positive class, and probabilities below 0.5, are mapped to the negative class). 








With respect to feature selection or model hyperparameter tuning, any of these procedures are carried out within the corresponding training set on the iterative procedure of cross-validation. For the specific case of the SVM classifier, a specific study was performed on a random subset of the data to determine a pseudo-optimal value for the $\gamma$ hyperparameter of the RBF kernel ($0.015$). This was necessary because of the sensitivity of SVMs, where slight changes in the value of this hyperparameter can result in very different results. For the remaining hyperparameters of the classifiers, pseudo-optimal values were found through a grid search, and the best settings were used to train the final models and compute the performance metrics included in Section~\ref{sec:Results}. On the other hand, the criteria chosen to determine which features are retained by the LASSO model were to keep all features whose associated $\beta_j$ coefficient had an absolute value equal to or greater than $10^{-4}$. Similarly, for the IG and RF feature selection methods, only those features with a coefficient (i.e., gain value or feature importance, respectively) higher than the arithmetic mean of all coefficients were retained to train any classifier further.
\section {Results}
\label{sec:Results}

Table~\ref{tab:all_results} presents the quantitative results obtained from the extensive analysis conducted in this study. This allows the comparison of all the different settings tested, starting from developing a classifier over the full dataset (i.e., with no use of any feature selection method), which is shown in the first block of rows within the table. It also includes the settings where all classifiers are trained after performing feature selection with the IG, LASSO, and RF algorithms, as shown in the second, third, and fourth rows of the table. The results include the average performance obtained by performing 10-fold cross-validation and are presented in terms of accuracy, precision, recall, F$_1$-Score, geometric mean, and area under the ROC curve. The resulting standard deviation in the 10 folds is included along with the different results obtained. The average number of features used to train the classifiers in each setting is shown in the table for further analysis and discussion.
After applying the Wilcoxon test to compare the significance of using ensemble models versus baseline classifiers to identify web traffic attacks, and for each feature selection setting analyzed, executions with a significant difference are marked with a star symbol (*).
The hyperparametric configuration of the different classifiers was as follows: for kNN, k=10; for SVM, a cost of 3000 was used, and for RF, the mtry was 50.

\begin{table*}[!ht]
    \centering
    \renewcommand{\arraystretch}{1.2}
    \caption{10-fold cross-validation average performance for the different classifiers and feature selection methods analyzed (FS=Feature Selection, Var=Number of input features, AUC=Area Under the ROC Curve).} 
    \label{tab:all_results}
\begin{tabular}{|c|c|c|c|c|c|c|c|c|}
    \toprule
    \textbf{FS} & \textbf{Classifier} & \textbf{Var} & \textbf{Accuracy} & \textbf{Precision} & \textbf{Recall} & \textbf{F1-Score} & \textbf{gmean} & \textbf{AUC} \\
    \midrule
    \multirow{5}{*}{-} 
    & LASSO & \multirow{5}{*}{78} & $0.718 \pm {\scriptstyle 0.01}$ & $0.831 \pm {\scriptstyle 0.02}$ & $0.773 \pm {\scriptstyle 0.01}$ & $0.801 \pm {\scriptstyle 0.01}$ & $0.801 \pm {\scriptstyle 0.01}$ & $0.797 \pm {\scriptstyle 0.01}$ \\
    & kNN  & & $0.718 \pm {\scriptstyle 0.01}$ & $0.817 \pm {\scriptstyle 0.01}$ & $0.781 \pm {\scriptstyle 0.01}$ & $0.798 \pm {\scriptstyle 0.01}$ & $0.799 \pm {\scriptstyle 0.01}$ & $0.798 \pm {\scriptstyle 0.01}$ \\
    & SVM  & & $0.736 \pm {\scriptstyle 0.16}$ & $0.756 \pm {\scriptstyle 0.20}$ & $0.819 \pm {\scriptstyle 0.13}$ & $0.784 \pm {\scriptstyle 0.18}$ & $0.786 \pm {\scriptstyle 0.01}$ & $0.866 \pm {\scriptstyle 0.17}$ \\
    & RF*  & & $0.952 \pm {\scriptstyle 0.01}$ & $0.964 \pm {\scriptstyle 0.01}$ & $0.965 \pm {\scriptstyle 0.01}$ & $0.965 \pm {\scriptstyle 0.01}$ & $0.965 \pm {\scriptstyle 0.01}$ & $0.983 \pm {\scriptstyle 0.01}$ \\
    & XGBoost* & & $0.964 \pm {\scriptstyle 0.01}$ & $0.970 \pm {\scriptstyle 0.01}$ & $0.976 \pm {\scriptstyle 0.01}$ & $0.973 \pm {\scriptstyle 0.01}$ & $0.973 \pm {\scriptstyle 0.01}$ & $0.989 \pm {\scriptstyle 0.01}$ \\
    \hline\noalign{\smallskip}
    \multirow{5}{*}{IG} 
    & LASSO & \multirow{5}{*}{29} & $0.715 \pm {\scriptstyle 0.01}$ & $0.841 \pm {\scriptstyle 0.02}$ & $0.765 \pm {\scriptstyle 0.01}$ & $0.801 \pm {\scriptstyle 0.01}$ & $0.802 \pm {\scriptstyle 0.01}$ & $0.789 \pm {\scriptstyle 0.01}$ \\
    & kNN  & & $0.716 \pm {\scriptstyle 0.01}$ & $0.824 \pm {\scriptstyle 0.01}$ & $0.774 \pm {\scriptstyle 0.01}$ & $0.798 \pm {\scriptstyle 0.01}$ & $0.799 \pm {\scriptstyle 0.01}$ & $0.793 \pm {\scriptstyle 0.01}$ \\
    & SVM & & $0.704 \pm {\scriptstyle 0.16}$ & $0.756 \pm {\scriptstyle 0.20}$ & $0.778 \pm {\scriptstyle 0.13}$ & $0.763 \pm {\scriptstyle 0.18}$ & $0.765 \pm {\scriptstyle 0.01}$ & $0.828 \pm {\scriptstyle 0.17}$ \\
    & RF* & & $0.948 \pm {\scriptstyle 0.01}$ & $0.957 \pm {\scriptstyle 0.01}$ & $0.966 \pm {\scriptstyle 0.01}$ & $0.961 \pm {\scriptstyle 0.01}$ & $0.961 \pm {\scriptstyle 0.01}$ & $0.973 \pm {\scriptstyle 0.01}$ \\
    & XGBoost* & & $0.957 \pm {\scriptstyle 0.01}$ & $0.973 \pm {\scriptstyle 0.01}$ & $0.965 \pm {\scriptstyle 0.01}$ & $0.969 \pm {\scriptstyle 0.01}$ & $0.969 \pm {\scriptstyle 0.01}$ & $0.987 \pm {\scriptstyle 0.01}$ \\
    \hline\noalign{\smallskip}
    \multirow{5}{*}{LASSO} 
    & LASSO & \multirow{5}{*}{35.1} & $0.716 \pm {\scriptstyle 0.01}$ & $0.831 \pm {\scriptstyle 0.01}$ & $0.771 \pm {\scriptstyle 0.01}$ & $0.800 \pm {\scriptstyle 0.01}$ & $0.801 \pm {\scriptstyle 0.01}$ & $0.796 \pm {\scriptstyle 0.01}$ \\
    & kNN  & & $0.719 \pm {\scriptstyle 0.01}$ & $0.822 \pm {\scriptstyle 0.01}$ & $0.778 \pm {\scriptstyle 0.01}$ & $0.800 \pm {\scriptstyle 0.01}$ & $0.800 \pm {\scriptstyle 0.01}$ & $0.797 \pm {\scriptstyle 0.01}$ \\
    & SVM & & $0.700 \pm {\scriptstyle 0.16}$ & $0.758 \pm {\scriptstyle 0.20}$ & $0.772 \pm {\scriptstyle 0.13}$ & $0.761 \pm {\scriptstyle 0.18}$ & $0.763 \pm {\scriptstyle 0.01}$ & $0.829 \pm {\scriptstyle 0.17}$ \\
    & RF* & & $0.963 \pm {\scriptstyle 0.01}$ & $0.971 \pm {\scriptstyle 0.01}$ & $0.975 \pm {\scriptstyle 0.01}$ & $0.973 \pm {\scriptstyle 0.01}$ & $0.973 \pm {\scriptstyle 0.01}$ & $0.986 \pm {\scriptstyle 0.01}$ \\
    & XGBoost* & & $0.962 \pm {\scriptstyle 0.01}$ & $0.968 \pm {\scriptstyle 0.01}$ & $0.976 \pm {\scriptstyle 0.01}$ & $0.972 \pm {\scriptstyle 0.01}$ & $0.972 \pm {\scriptstyle 0.01}$ & $0.987 \pm {\scriptstyle 0.01}$ \\
    \hline\noalign{\smallskip}
    \multirow{5}{*}{RF} 
    & LASSO & \multirow{5}{*}{26.9} & $0.706 \pm {\scriptstyle 0.03}$ & $0.872 \pm {\scriptstyle 0.05}$ & $0.747 \pm {\scriptstyle 0.05}$ & $0.802 \pm {\scriptstyle 0.01}$ & $0.806 \pm {\scriptstyle 0.03}$ & $0.782 \pm {\scriptstyle 0.01}$ \\
    & kNN & & $0.720 \pm {\scriptstyle 0.01}$ & $0.803 \pm {\scriptstyle 0.01}$ & $0.790 \pm {\scriptstyle 0.01}$ & $0.797 \pm {\scriptstyle 0.01}$ & $0.797 \pm {\scriptstyle 0.01}$ & $0.798 \pm {\scriptstyle 0.01}$ \\
    & SVM & & $0.774 \pm {\scriptstyle 0.18}$ & $0.819 \pm {\scriptstyle 0.21}$ & $0.845 \pm {\scriptstyle 0.17}$ & $0.832 \pm {\scriptstyle 0.20}$ & $0.832 \pm {\scriptstyle 0.02}$ & $0.846 \pm {\scriptstyle 0.19}$ \\
    & RF* & & $0.965 \pm {\scriptstyle 0.01}$ & $0.970 \pm {\scriptstyle 0.01}$ & $0.977 \pm {\scriptstyle 0.01}$ & $0.974 \pm {\scriptstyle 0.01}$ & $0.974 \pm {\scriptstyle 0.01}$ & $0.987 \pm {\scriptstyle 0.01}$ \\
    & XGBoost* & & $0.963 \pm {\scriptstyle 0.01}$ & $0.971 \pm {\scriptstyle 0.01}$ & $0.974 \pm {\scriptstyle 0.01}$ & $0.973 \pm {\scriptstyle 0.01}$ & $0.973 \pm {\scriptstyle 0.01}$ & $0.988 \pm {\scriptstyle 0.01}$ \\
    \bottomrule
\end{tabular}
\end{table*}

\begin{figure}[!ht]
    \centering
    \includegraphics[width=0.8\linewidth]{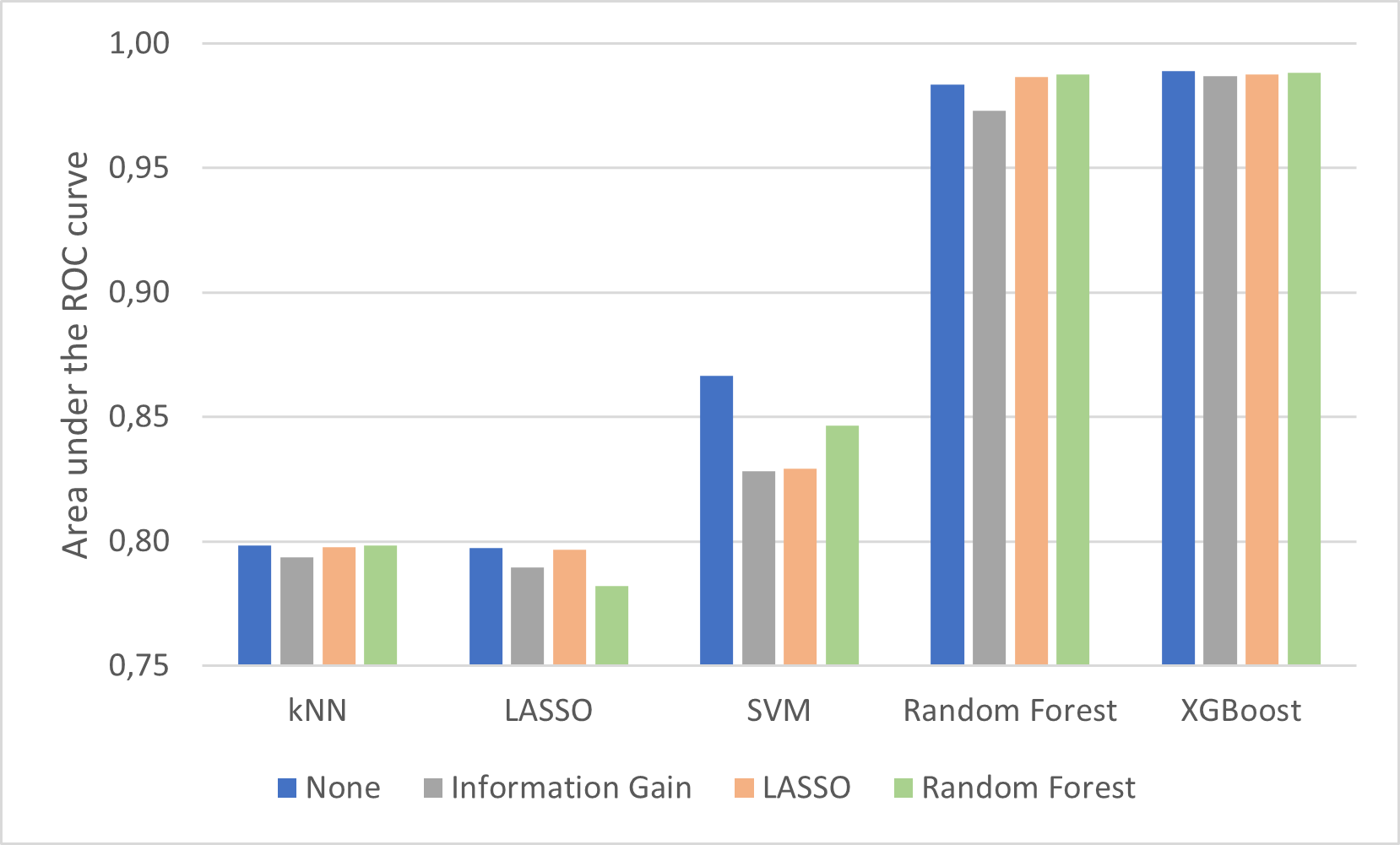}
    \caption{Average AUC performance of all classifiers and feature selection methods analyzed.}
    \label{fig:BarPlotSinSel}
\end{figure}

Because the problem addressed in this work is an imbalanced binary classification task, we first focus on the AUC metric achieved by different settings. The average AUC performance is shown in Figure~\ref{fig:BarPlotSinSel}. Concerning the different classifiers tested in this study, it is evident that ensemble methods, either RF or XGBoost, significantly outperform any of the three baseline methods considered (kNN, LASSO, and SVM). If we consider the best AUC achieved by the kNN, LASSO, and SVM classifiers ($0.798$, $0.797$, $0,866$, respectively) on average, the best performance of the baseline models for identifying web attacks is $0.82$. However, if the same approach is followed for the ensemble classifiers ($0.987$ and $0.989$ for RF and XGBoost, respectively), on average, the best performance increases up to $0.988$. In other words, the use of ensemble methods allows achieving a positive impact on the predictive performance of classifiers, which is approximately $20\%$ more accurate than baseline models.

Regarding the feature selection methods considered in this study, two different results were identified. First, there seems to be no positive impact on the predictive performance when applying feature selection prior to training any classifier. For both the baseline and ensemble models analyzed, the best AUC achieved was the setting where no feature selection was performed ($0.798$, $0.797$, $0.866$, $0.989$ for kNN, LASSO, SVM, and XGBoost, respectively), except for the RF model, which achieved an AUC of $0.987$ after filtering features through the importance assigned by itself during the training process. However, for the latter, the use of the no feature selection method achieves an AUC of $0.983$ which is not significantly worse than the best AUC. Second, if one averages the standard deviation of the AUC achieved by each classifier with the four different feature selection options, it is important to highlight that baseline models have more than twice the average standard deviation than ensemble models, which turns out that baseline models are much more sensitive and more likely to achieve better performance results than ensemble models.

Although feature selection methods have not been shown to be relevant in this context (in a positive manner), it is also worth noting that they are capable of achieving similar performance to classifiers trained over the full dataset, but using much fewer input features. Table~\ref{tab:all_results} shows that the IG, LASSO, and RF algorithms retain an average of $29$, $35.1$ and $26.9$ features within the cross-validation strategy, respectively, in contrast to the $78$ input features that are present in the dataset. This means that on average, feature selection methods achieve similar AUC performance when using less than half of the initial input features. Suppose one joins this result with those described in the previous paragraph. In this case, it is important to point out that feature selection methods, particularly when ensemble classifiers are used, could potentially become relevant in a context where some difficulties are encountered to collect or access all the required input features for the models.

\begin{figure}[!ht]
\centering
	\subfloat[Comparison of classifiers]{\includegraphics[width=3in]{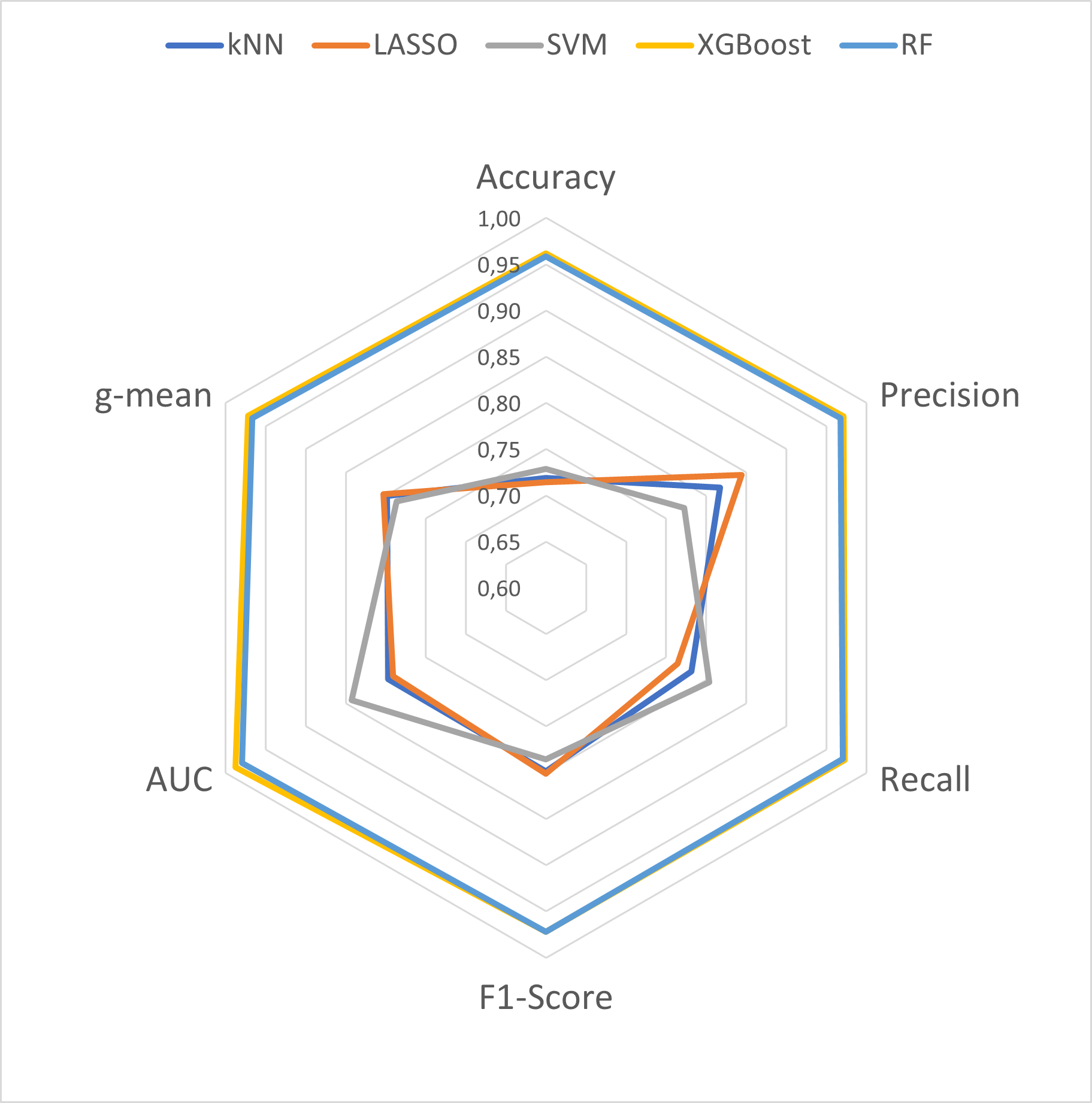}}
    \hspace{2mm}
    \color{lightgray} \rulesep \color{black}
    \hspace{2mm}
	\subfloat[Comparison of feature selection methods]{\includegraphics[width=3in]{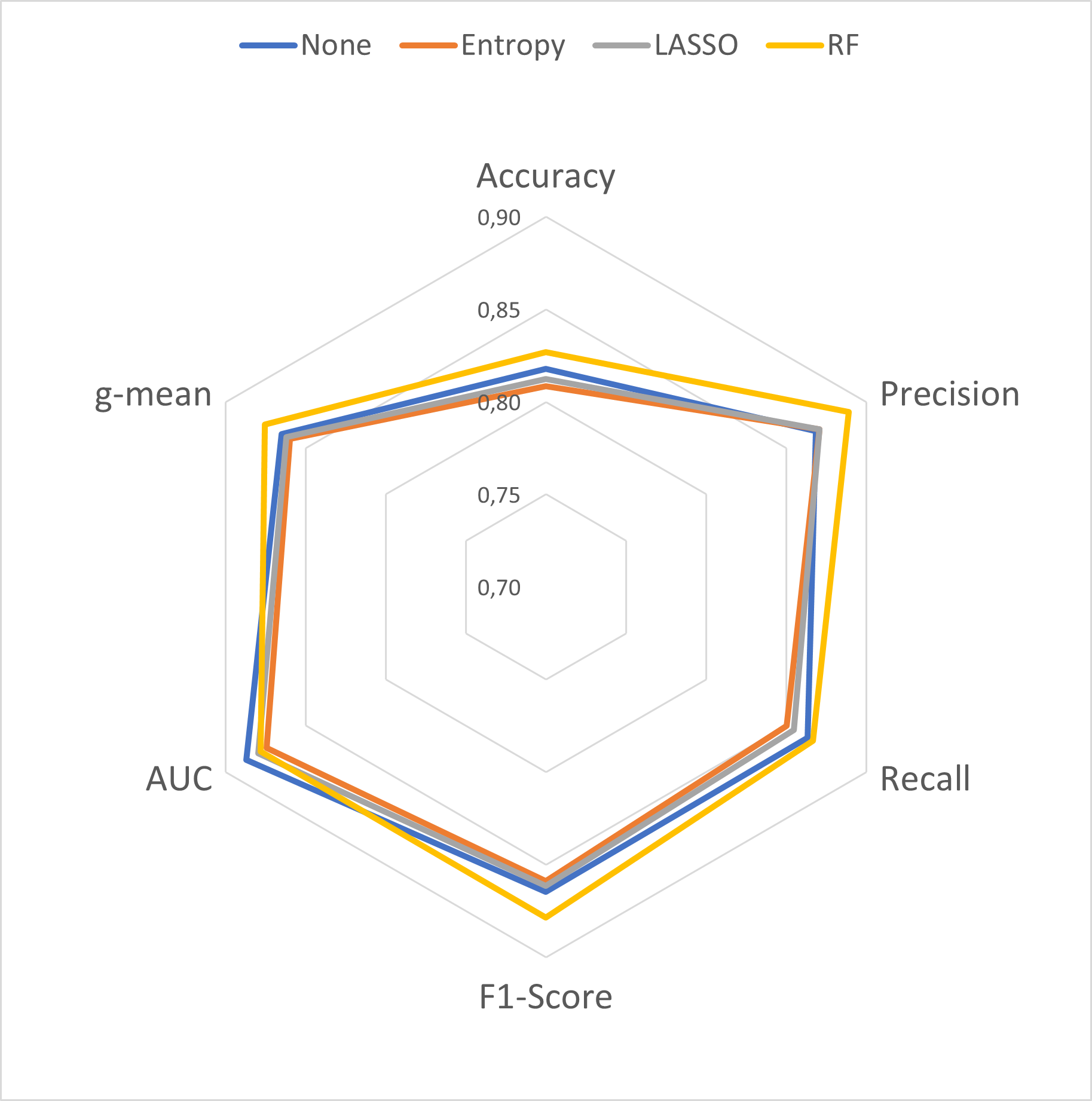}}
\caption{Average performance of all metrics from different perspectives.}
\label{fig:RadarPlots}
\end{figure}

To complement the analysis of the performance of the different settings tested in this work, Figure~\ref{fig:RadarPlots} shows the average performance in terms of all the different metrics considered in this work, either from the perspective of the classifiers (Fig. ~\ref{fig:RadarPlots}a)] or feature-selection methods [Fig. ~\ref{fig:RadarPlots}b). In particular, the F$_1$-Score measure is also a good indicator of the predictive performance of classifiers because it considers the accuracy of the positive predictions made by the model (precision) and the ability of the model to capture all positive instances (recall). In this sense, Figure~\ref{fig:RadarPlots}a illustrates how XGBoost obtains, on average, the highest F$_1$-Score measure of $0.9721$ regardless of the previously employed feature selection method. The RF algorithm is illustrated in Fig. ~\ref{fig:RadarPlots}b as the feature selection method, which, on average, obtained the highest F$_1$-Score of $0.8694$ independent of the classifier used. This confirms the results previously presented in this study, in which XGBoost turned out to be the best performing classifier; in the case of requiring any feature selection method, the RF algorithm would be the best choice to obtain similar performance results with less than half of the input features.

\section {Conclusions and future work}
\label{sec:ConcAndFW}

This paper presents an extensive analysis of the development of highly accurate predictive models to discriminate attacks from normal traffic behavior. For this purpose, the CSIC2010 v2 dataset was employed, which provides a sufficiently large dataset containing samples of normal and anomalous traffic within a web application context in which users can purchase any offered product. This study analyzed the predictive performance of five different classifiers distributed among the baseline classifiers (kNN, LASSO, and SVM with an RBF kernel) and ensemble classifiers (RF and XGBoost). Furthermore, it analyzed the impact of employing three well-known feature selection methods prior to fitting the classifiers: IG, LASSO, and RF.

This study has shown an approach for extracting relevant features from rich information encoded in HTTP traces, such as payloads or URL. Although this work focused on a specific web application, the same approach could be employed on any other similar web application to build an ad hoc dataset and develop an accurate predictive model specialized in that specific context. Despite the existence of newer and more secure web applications in the market, in which payloads may nowadays be encrypted among other security measures, this study is still potentially interesting to many small- or medium-sized companies that have been efficiently working with their web application for many years and cannot afford the costs of changing or updating their systems (e.g., monetary and operational). In this sense, this study provides a simple yet effective way of implementing an automatic attack identification system to be deployed at the back end of their web application and, consequently, implementing certain security rules to be triggered when attacks are detected.

Furthermore, the results obtained in the analysis showed that using ensemble methods (RF or XGBoost) can have a positive impact on the predictive performance of classifiers. Specifically, ensemble methods have achieved an average AUC that is approximately $20\%$ higher than the baseline models (kNN, LASSO, or SVM); XGBoost outperformed all the other classifiers with an average AUC of $0.989$. In addition, it has been shown that the use of feature selection methods together with ensemble classifiers turned into a less sensitive model developing approach than using them with baseline models because the variability of the predictive performance in the trained classifiers is, on average, less than half when ensemble models are used. This result is potentially useful and interesting for scenarios in which collection and access to all required features is difficult or even impossible. While ensemble models, such as RF and XGBoost, demonstrate significant improvements in predictive performance, these gains come with trade-offs, including increased computational complexity and longer training times.

Altogether, these results encourage us to continue working on this research topic in the near future. For instance, strategies to optimize ensemble methods for time efficiency and evaluate their feasibility in real-time applications can be further explored. Moreover, the methodology proposed to extract features from relevant fields encoded in HTTP traces and the findings described in this study can be validated using different web applications. Furthermore, it would be interesting to extend this type of intrusion or attack detection system to different contexts aside from web applications, such as IoT devices, sensors, and network communications. Finally, further work could explore HTTPS traffic analysis, the application of the methodology to more recent and diverse open datasets, its potential use as an Intrusion Detection System (IDS), and the challenges of handling unbalanced datasets or live network traffic capture in real-world scenarios.

\section*{Acknowledgements}
This research is part of the AI4SECIoT Strategic Project  (``Artificial Intelligence for Securing IoT Devices''), under the collaboration agreement signed between the National Cybersecurity Institute (INCIBE) and the University of Burgos. This initiative is carried out within the framework of the Recovery, Transformation and Resilience Plan funds, financed by the European Union (Next Generation). In addition, this work resulted from the collaboration established by Daniel Urda during his postdoc internship at the University of Tartu.

\bibliographystyle{splncs04}
\bibliography{bibliography}

\end{document}